# Crystal Growth and Doping Control of HgBa$_2$CuO$_{4+\delta}$, the Model Compound for High-$T_c$ Superconductors

Anaëlle Legros,[†,‡] Bastien Loret,[†,‡,#] Anne Forget,[†] Patrick Bonnaillie,[§] Gaston Collin,[∥] Pierre Thuéry,[⊥] Alain Sacuto,[#] and Dorothée Colson[†,*]

[†]SPEC, CEA, CNRS UMR 3680, [§]SRMP, CEA, DMN, and [⊥]NIMBE, CEA, CNRS, Université Paris-Saclay, 91191 Gif sur Yvette Cedex, France

[‡]Department of Physics, University of Sherbrooke, Sherbrooke, QC J1K 2R1, Canada

[∥]LPS, CNRS UMR 8502, Université Paris-Saclay, 91405 Orsay, France

[#]Laboratoire Matériaux et Phénomènes Quantiques, 10 rue A. Domon et L. Duquet, 75205 Paris Cedex 13, France

**ABSTRACT:** A new method to grow very high quality single crystals of the superconducting HgBa$_2$CuO$_{4+\delta}$ mercury cuprates is reported. The single crystals are platelet-shaped, with surfaces of high optical quality and good crystallographic properties. Annealing enables optimization of $T_c$ up to $T_c^{max}$ = 94 K. With adequate treatment, the doping level of the crystalline samples can be finely controlled in a wide under- and over-doped range. Preliminary structural characterization from single crystal X-ray diffraction data is given for different doping levels. The role of added gold on the doping is also investigated. The signature of under- and over-doping for both pure and gold-substituted crystals has been identified from micro-Raman spectroscopy measurements.

Since the discovery of the high-$T_c$ superconductors (HTSC) in 1986,[1] mercury cuprates HgBa$_2$Ca$_{n–1}$Cu$_n$O$_{2n+2+\delta}$, where $n$ is the number of CuO$_2$ layers, play a peculiar role as being the structurally simplest cuprates with the highest $T_c$ values, up to 133 K (160 K under 30 GPa) for HgBa$_2$Ca$_2$Cu$_3$O$_{8+\delta}$ (Hg-1223).[2–5] The HgBa$_2$CuO$_{4+\delta}$ compound (Hg-1201) with only one CuO$_2$ plane per tetragonal unit cell (space group $P4/mmm$, Figure 1, left), and the highest $T_c$ value of all single-layer cuprates, is an ideal candidate to clarify the relationship between the crystallographic structure and the electronic properties, and to improve the comprehension of high-$T_c$ superconductivity mechanisms so as to reach still higher $T_c$ values. Unfortunately, few single crystal studies have been performed, notwithstanding some progress in Hg-1201 crystal growth in the last decades.[6–13] Here we report a new method to grow single crystals of Hg-1201 cuprate using self-flux techniques. The surfaces of the crystals synthesized are extremely clean, with high optical quality and good crystallographic properties (Figure 1, right). Optimization of $T_c$, up to $T_c^{max}$ (94 K), is possible through annealing, a wide range of doping levels in the under- and over-doped regimes being accessible through adequate heat treatments as previously described on powders.[14-16] Crystals elaborated by the same method have previously been used in Raman studies.[17,18]

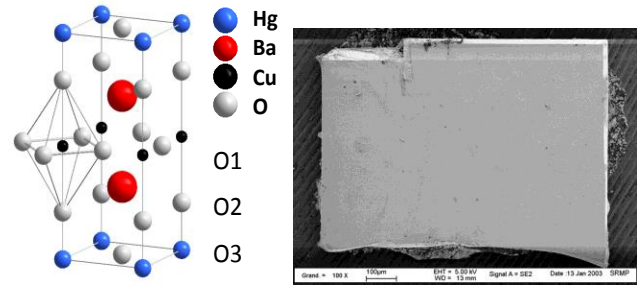

**Figure 1.** Schematic representation of the crystal structure of tetragonal HgBa$_2$CuO$_{4+\delta}$ (left) and scanning electron microscopy (SEM) view of a single crystal of HgBa$_2$CuO$_{4+\delta}$ (right). The doping process consists in inserting oxygen on the O3 site, which is thus partially occupied (δ occupancy). The surfaces have high optical quality, and the crystallographic axes are easily identified ($a$ axis along the edges, $c$ axis perpendicular to the platelet).

As already signalled in our recent communication about Hg-1223,[19] these mercury compounds are very attractive, but they are complex oxides and their synthesis remains a challenge, in particular due to the toxicity of mercury oxide HgO and the necessity to control the volatility of its decomposition products. The present synthesis procedure allows the synthesis of large pure crystals and the easy separation of the sample from the flux, thus providing high quality crystals with clean surfaces. It also has the advantage to be less expensive and more simple to operate than the technique using gold foil.[6,7,8]

A low melting region in the pseudo-ternary HgO–BaO–CuO phase diagram was determined, and crystal growth was achieved in a melt with an excess of BaO and CuO (see Experimental Section, Supporting Information). In our experimental conditions, the interesting domain is bounded by BaCuO$_2$, BaO and Hg-1201 (Figure S1, Supporting Information). The main compositions along the Hg-1201–Ba$_2$CuO$_3$ line have been investigated. In order to determine a suitable growth temperature for the compositions labelled A to K, samples were heated to temperatures between 950 and 1000 °C and then cooled down at 5–20 °C/h. The experiments have revealed that a step at 750 °C ensures complete reaction by minimizing the soaking time. The composition most favourable to the growth of crystals was that labelled K, for which 16.2 mol% HgO and 59:24.8 mol% BaO:CuO were mixed. After the growth stage, most crystals are usually found at the bottom of the crucible in the frozen flux, from which they are then mechanically separated; some large crystals may be found in the more favourable cases (Figure S2). Figure 1 shows one typical, extracted platelet-shaped crystal, with well-developed {001} faces and a size of 0.7 × 0.5 × 0.05 mm$^3$. With small cooling rates of 5 °C/h, thicknesses up to 0.3 mm have also been obtained.

A remarkable point is that the crystallographic axes of the crystalline platelets are easily identifiable (*a* axis along the edges, *c* axis perpendicular to the platelet largest face), as in the Hg-1223 phase.[16] The chemical analysis of as-grown crystals, performed with a scanning electron microscope equipped with an electron microprobe, revealed a slight under-stoichiometry of ~0.94 in mercury, the mercury content being homogeneous in each crystal. No flux component was detected.

In order to increase the doping level of the CuO$_2$ layers as in the Au-substituted Hg-1223 compound[20] and thus access more easily the over-doped regime, we have elaborated other crystals by substituting Hg by Au, through addition of gold powder among starting reagents. Chemical analyzes carried out on the single crystals thus obtained showed that the average composition is (Hg$_{0.93}$Au$_{0.04}$)Ba$_2$CuO$_z$. The distribution in gold is very homogeneous within each crystal and from one crystal to another (an average of 10 measurement points were made on the surface and in depth for each crystal). There is a very slight sub-stoichiometry (Hg + Au) ≈ 0.97, which is within the resolution limit of the device and indicates a good agreement with the expected formula.

The crystal structures of five samples of gold-free (**UD55**, **UD62**, **UD73**, **OP93** and **OD89**) and four samples of gold-substituted Hg-1201 (**UD72**, **UD88**, **OD72** and **OD65**) were refined from single crystal X-ray diffraction experiments (see Supporting Information for details). Atomic coordinates are given in Table S1, and crystal data and values of refined parameters in Tables S2 and S3. Although this structural work is still preliminary, some trends are apparent in the present results. In all cases, refinement of the occupancy parameter of mercury led to a much-improved refinement quality and, more importantly, a much more acceptable displacement parameter for this atom, abnormally large values being found for full occupancy.[6] These occupancies are in the range of 0.933(2)–0.9720(12) and, as a general trend, they decrease when the amount of oxygen in the sample increases, the lower value being for **OD89**. The occupancy parameter for atom O3, at (½ ½ 0), can only be obtained with a large uncertainty, due to the low associated electronic density and the presence of several heavy atoms in the structure. The corresponding electronic density appears however very clearly in the Fourier difference maps for the most oxygenated samples **OP93** and **OD89**. In the former, refinement proceeds smoothly, provided that O3 is refined isotropically (anisotropic refinement being highly unstable) and, since the refined displacement parameter of O3 is close to the equivalent isotropic displacement parameter of O1, this has been used as a restraint for all other samples, for which free refinement gave unreasonable values. When taking into account the large standard deviations, the refined O3 occupancies in the series of gold-free samples agree roughly with the values calculated from the relation between $T_c/T_c^{max}$ and the number of carriers in the CuO$_2$ plane[21] (Table S2). Except for the as-grown crystal **UD73**, we can note a decrease of the *a* parameter with the doping level, which is common in HTSC compounds (Table S2).[8,16] For all gold-substituted samples, no significant electronic density is found at (½ ½ 0), suggesting the complete absence of oxygen at this site.

Magnetic measurements show that all the as-grown samples are under-doped with a $T_c$ (midpoint) around 70 K (**UD73**) and a fairly sharp transition ($\Delta T_c < 4$ K) as shown in Figure 2. In order to increase $T_c$ to $T_c^{max}$, single crystals have been annealed under molecular oxygen flux at 325 °C. The crystals were placed in advance in a mixture of HgO, BaO and CuO oxides to prevent a departure of mercury from the samples during the 10 days heat treatment.

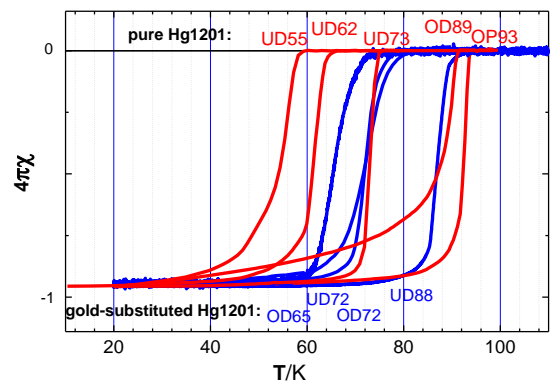

**Figure 2.** Temperature-dependence of the normalized magnetic susceptibility of under-doped (UD), optimal (OP) and over-doped (OD) crystals of pure (red) and gold-substituted (blue) HgBa$_2$CuO$_{4+\delta}$.

After annealing, $T_c$ increases up to 93 K (**OP93**) with a narrow transition width of 5 K confirming the good homogeneity of the oxygen content in the sample. In order to access the under-doped region, we also developed a procedure of annealing under vacuum. As-grown crystals have been annealed under a vacuum of 6.0 10$^{-7}$ mbar at 400 °C during 4 days, or at 450 °C during 2 days. These thermal treatments result in a decrease of $T_c$ from 70 K to 62 K (**UD62**) and 55 K (**UD55**), respectively, as shown in Figure 2. Over-doping the crystals under oxygen high pressure (80 bars) at 300 and then 280 °C during 3 days gives sample **OD89**, with a $T_c$ of 89 K.

Magnetic transitions of $Hg_{1-x}Au_xBa_2CuO_{4+\delta}$ gold-substituted crystals (Au-Hg-1201 hereafter) have also been determined. The Au-Hg-1201 as-grown crystals have a $T_c$ of 88 K (**UD88**) which is much higher than the $T_c$ value for pure Hg-1201 ($T_c$ = 72 K) elaborated in the same conditions. Several crystals have been annealed under a vacuum of 6.0 10$^{-7}$ mbar at 400 °C during 5 days. This treatment considerably decreases $T_c$, from 88 K to 72 K (**UD72**). Inversely, over-doping and decrease of $T_c$ from 88 K to 72 K (**OD72**) follow treatment under oxygen flux at 325 °C during 8 days. To achieve an even larger over-doping of the Au-Hg-1201 samples, some as-grown crystals have been annealed under a vacuum of 6.0 10$^{-7}$ mbar at 400 °C during 4 days and then under oxygen flux at 300 °C for 10 days. In this case $T_c$ decreases, from 88 K (**UD88**) in the under-doped regime to 65 K (**OD65**) in the over-doped regime.

All crystals have been characterized by micro Raman spectroscopy which reveals for the first time the specific Raman features of the under-doped, optimally and over-doped pure or gold-substituted Hg-1201 compounds. $HgBa_2CuO_{4+\delta}$ has a very simple structure with a single $CuO_2$ plane as a mirror plane. As a consequence, Raman spectroscopy is blind to the normal vibrational modes related to the $CuO_2$ plane for selection rule reasons.[22-24] Solely, the $A_{1g}$ vibrational mode related to the vertical motion of the apical oxygen O2 is observable in the spectra of pure Hg-1201 (Fig. 3a). The O2 mode is located at 597 cm$^{-1}$ for **UD55**. As the oxygen doping increases, the O2 mode softens in frequency and becomes broader. At the optimal doping (**OP93**) it is located at 592 cm$^{-1}$ and it is accompanied by a new feature located around 570 cm$^{-1}$. We interpret the softening and the broadening of the O2 mode as the consequence of the enhancement of the charge transfer between the Hg–O and $CuO_2$ plane induced by the oxygen atoms insertion on the O3 site during the doping process. In this scenario, the new feature corresponds to the extra oxygen vibrational mode related to the O3 site brought by doping. Remarkably, the O2 vibrational mode stiffens in frequency in the gold substituted crystals with respect to the free gold ones, This is shown in Fig. 3c where a shift of 3 cm$^{-1}$ toward high frequencies is detected for the O2 mode in Au-Hg-1201 with respect to the pristine one for nearly the same doping level (**UD73**). Importantly, no extra feature around 570 cm$^{-1}$ related to the O3 site occupancy is detected in Au-Hg-1201 crystals even in the over-doped regime (Fig. 3b). This suggests that the Au substitution in Hg-1201 favors the hole doping in the $CuO_2$ plane without necessarily introducing a large amount of oxygen atoms on the O3 site of the Hg–O plane. As a consequence, a weaker amount of oxygen content is required to reach the optimal doping level in the Au-Hg-1201 crystals than in the pure ones.

In this communication we propose an original procedure for the elaboration of high quality submillimetre platelet crystals of superconducting $HgBa_2CuO_{4+\delta}$. The remarkable points are that the surfaces of the crystals are extremely clean with high optical quality and the crystallographic axes are easily identifiable. An annealing method to finely control the oxygen content of samples in the under- and over-doped regimes is also described. Raman measurements reveal the interplay between the Hg/Au substitution and the oxygen insertion in the hole doping process of the Hg-1201 crystals, in perfect agreement with the X-ray diffraction measurements. It is notable that this crystal growth method can be applied to other compounds for which there exists no phase diagram and/or the growth in air is forbidden because of volatile components.

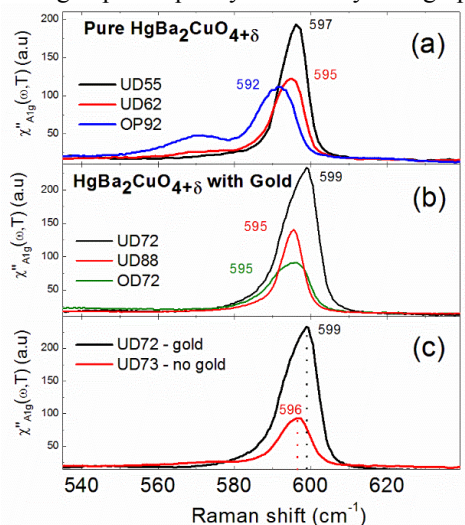

**Figure 3.** Raman spectra of pristine and gold-substituted Hg-1201 single crystals for several doping levels.

## ASSOCIATED CONTENT

**Supporting Information**
The supporting Information is available free of charge on the ACS Publications website at DOI:
Experimental details, additional figures, and tables (PDF)
**Accession Codes**
CSD 1868253−1868261 contain the supplementary crystallographic data for this paper. These data can be obtained free of charge via www.ccdc.cam.ac.uk/data_request/cif, or by emailing data_request@ccdc.cam.ac.uk, or by contacting The Cambridge Crystallographic Data Centre, 12 Union Road, Cambridge CB2 1EZ, UK; fax: +44 1223 336033.

## AUTHOR INFORMATION
**Corresponding Author**
*E-mail: dorothee.colson@cea.fr (D.C.).
**ORCID**
Dorothée Colson: 0000-0002-5213-5375

**Notes**

The authors declare no competing financial interest.

**REFERENCES**

(1) Bednorz, J.G.; Müller, K. Possible High-Tc superconductivity.in the Ba-La-Cu-O system. *Zeit. Phys. B* **1986**, *64*, 189–193.

(2) Putilin, S. N.; Antipov, E. V.; Chmaissem, O.; Marezio, M. Superconductivity at 94 K in $HgBa_2CuO_{4+\delta}$. *Nature* **1993**, *362*, 226–228.

(3) Schilling, A.; Cantoni, M.; Guo, J. D.; Ott, H. R. Superconductivity above 130 K in the Hg–Ba–Ca–Cu–O system. *Nature* **1993**, *363*, 56–58.

(4) Chu, C. W.; Gao, L.; Chen, F.; Huang, Z. J.; Meng, R. L.; Xue, Y. Y. Superconductivity above 150 K in $HgBa_2Ca_2Cu_3O_{8+\delta}$ at high pressures. *Nature* **1993**, *365*, 323–325.

(5) Nunez-Regueiro, M.; Tholence, J. L.; Antipov, E. V.; Capponi, J. J.; Marezio, M. Pressure-induced enhancement of Tc above 150 k in Hg-1223. *Science* **1993**, *262*, 97–99.

(6) Bertinotti, A.; Viallet, V.; Colson, D.; Marucco, J. F.; Hammann, J.; Le Bras, G.; Forget, A. Synthesis, crystal structure and magnetic properties of superconducting single crystals of $HgBa_2CuO_{4+\delta}$. *Physica C* **1996**, *268*, 257–265.

(7) Bertinotti, A.; Colson, D.; Marucco, J.-F.; Viallet, V.; Le Bras, G.; Fruchter, L.; Marcenat, C.; Carrington, A.; Hammann, J. Single crystals of mercury based cuprates: growth, structure and physical properties., Studies of High Tc superconductors, Ed. Narlikar, Nova Science Publisher (NY) **1997**, *23*, 27–85.

(8) Viallet-Guillen, V. Synthèse, études structurales et physico-chimiques de monocristaux d'oxydes supraconducteurs $HgBa_2Ca_{n-1}Cu_nO_{2n+2+\delta}$. PhD Thesis, Orsay, **1998**.

(9) Bordet, P.; Duc, F.; Le Floch, S.; Capponi, J. J.; Alexandre, E.; Rosa-Nunes, S.; Antipov, E. V. Single crystal X-ray diffraction study of the $HgBa_2CuO_{4+\delta}$ superconducting compound. *Physica C* **1996**, *271*, 189–196.

(10) Pissas, M.; Billon, B.; Charalambous, M.; Chaussy, J.; Le Floch, S.; Bordet, P.; Capponi, J. J. Single-crystal growth and characterization of the superconductor $HgBa_2CuO_{4+\delta}$. *Supercond. Sci. Techn.* **1997**, *10*, 598–604.

(11) Pelloquin, D.; Hardy, V.; Maignan, A.; Raveau, B., Single crystals of the 96 K superconductor $(Hg,Cu)Ba_2CuO_{4+\delta}$: growth, structure and magnetism. *Physica C* **1997**, *273*, 205–212.

(12) Pelloquin, D., Maignan, A; Guesdon, A.; Hardy, V.; Raveau, B. Single crystal study of the "1201" superconductor $Hg_{0.8}Bi_{0.2}Ba_2CuO_{4+\delta}$. *Physica C* **1996**, *265*, 5–12.

(13) Zhao, X.; Yu, G.; Cho, Y. C.; Chabot-Couture, G.; Barisic, N.; Bourges, P.; Kaneko, N.; Lu, L.; Motoyama, E. M.; Vajk., O. P.; Greven M. Crystal growth and characterization of the model high-temperature superconductor $HgBa_2CuO_{4+\delta}$. *Adv. Mater.* **2006**, *18*, 3243–3247.

(14) Marucco, J.-F.; Viallet V.; Bertinotti, A.; Colson, D.; Forget, A. Point defects, thermodynamic and superconducting properties of the non-stoichiometric $HgBa_2CuO_{4+\delta}$ phase, *Physica C* **1997**, *275*, 12–18.

(15) Yamamoto, A.; Hu, W. Z.; Izumi, F.; Tajima, S. Superconducting and structural properties of nearly carbonate-free $HgBa_2CuO_{4+\delta}$, *Physica C* **2001**, *351*, 329–340.

(16) Yamamoto, A.; Hu, W. Z.; Tajima, S. Thermoelectric power and resistivity of $HgBa_2CuO_{4+\delta}$ over a wide doping range. *Phys. Rev. B* **2001**, *63*, 024504.

(17) Loret, B.; Sakai, S.; Benhabib, S.; Gallais, Y.; Cazayous, M.; Measson, M. A.; Zhong, R. D.; Schneeloch, J.; Gu, G. D.; Forget, A.; Colson, D.; Paul, I.; Civelli, M.; Sacuto A. Vertical temperature-boundary of the pseudogap under the superconducting dome of the $Bi_2Sr_2CaCu_2O_{8+\delta}$ phase-diagram. *Phys. Rev. B* **2017**, *96*, 094525.

(18) Loret, B.; Sakai, S.; Gallais, Y.; Cazayous, M.; Méasson, M. A.; Forget, A.; Colson, D.; Civelli, M.; Sacuto A. Unconventional high-energy-State contribution to the Cooper pairing in the underdoped copper-oxides uperconductor $HgBa_2Ca_2Cu_3O_{8+\delta}$. *Phys. Rev. Lett.* **2016**, *116*, 197001.

(19) Loret, B.; Forget, A.; Moussy, J. B.; Poissonnet, S.; Bonnaillie, P.; Collin, G.; Thuéry, P.; Sacuto, A.; Colson, D. Crystal Growth and Characterization of $HgBa_2Ca_2Cu_3O_{8+\delta}$ Superconductors with the Highest Critical Temperature at Ambient Pressure, *Inorg. Chem.* **2017**, *56*, 9396–9399.

(20) Bordet, P.; Le Floch, S.; Capponi, J. J.; Chaillout C.; Gorius, M. F.; Marezio, M.; Tholence, J. L.; Radaelli, P. G. Gold substitution in mercury cuprate superconductors. *Physica C* **1996**, *262*, 151–158.

(21) Presland, M.R.; Tallon, J.L.; Buckley, R.G.; Liu, R.S.; Flower, N.E. General trends in oxygen stoichiometry effects on $T_c$ in Bi and Tl superconductors. *Physica C* **1991**, *176*, 95–105.

(22) G. Burns Introduction to group theory with application, Materials sciences series, IBM Thomas J. Watson Research Center: Yorktown Heights, New York, 1977.

(23) Guyard, W.; Cazayous, M.; Sacuto, A; Colson, D. Experimental evidences for a strong coupling between electrons and the apical oxygen phonon of $HgBa_2CuO_{4+\delta}$, *Physica C* **2007**, *460-462*, 380–381.

(24) Krantz, M. C.; Thomsen, C.; Mattausch, Hj.; Cardona, M., Raman Active Phonons and Mode Softening in Superconducting $HgBa_2CuO_{4+\delta}$, *Phys. Rev. B* **1994**, *50*, 1165–1170.

# Supporting Information

# Crystal Growth and Doping Control of HgBa$_2$CuO$_{4+\delta}$, the Model Compound for High-Tc Superconductors

Anaëlle Legros, Bastien Loret, Anne Forget, Patrick Bonnaillie, Gaston Collin, Pierre Thuéry, Alain Sacuto and Dorothée Colson

**Experimental Section**

*Crystal growth:* The starting mixtures for crystal growth were prepared using HgO (99%), BaO (99.5%), and CuO (99.995%) oxides, as well as gold powder (99.95%) in some experiments. All manipulations of oxides were carried out in an Ar-gas-filled glove box. The reagents were weighted, mixed and ground, giving powders (2 g) with the nominal compositions indicated in Figure S1. The best results were obtained by using the composition labelled K. For samples with gold, 2 g of oxides with the composition K were mixed with 250 mg of gold powder.

The mixtures were directly placed in alumina crucibles and sealed in evacuated quartz tubes (13 mm in diameter, 70 mm in length). For the sake of safety, the tubes were enclosed in stainless steel containers. Heat treatments were carried out in vertical tubular furnaces. Crystal growth was most favoured at temperatures close to 995 °C. Typical growth conditions were as follows: the sample was heated at 750 °C for 10 h, then further heated to 995 °C at 100 °C/h and soaked for 1 h. The temperature was afterwards lowered to 800 °C at a rate of 5-20 °C/h, and the furnace was cooled down to room temperature at 200 °C/h.

**Figure S1.** Phase diagram for HgO-BaO-CuO in air. Our study has identified a region of low melting compositions for application of the nonstoichiometric melt method. A composition close to "K" is favorable for synthesizing Hg-1201 crystals.

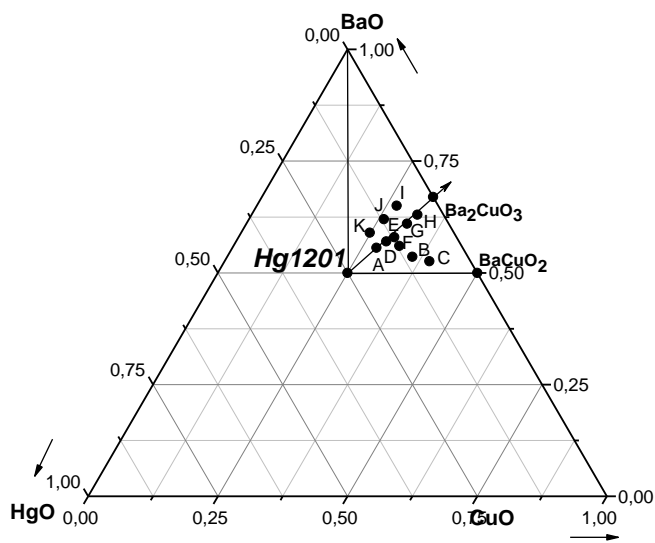

**Figure S2.** Single crystal of HgBa$_2$CuO$_{4+\delta}$ of 4 × 5 mm$^2$ formed at the bottom of the crucible and surrounded by frozen flux (SEM).

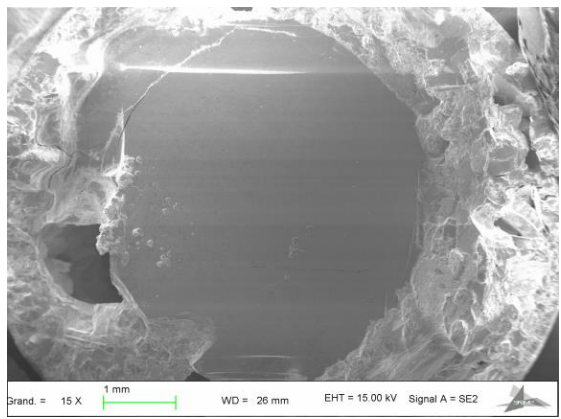

*Microprobe analysis:*

The chemical composition of the crystals has been determined at several locations on the surface with a Camebax SX50 electron microprobe. The analyzed volume is of a few (μm)$^3$. Measurements in depth along a line of 110 μm of a cut and polished crystal (see below, the electronic micrograph) have also been performed. The sum of the atomic percentages of Hg, Ba, Cu and O equals 100. The atomic percentage for the Al atom is below the limit of detection. These quantitative data thus show no contamination inside the crystal by the alumina crucible and no carbon as impurity on the mercury site.

**Figure S3.** Composition profile for the elements Hg, Ba, Cu, Al and O inside a Hg-1201 crystal, as indicated on the scanning electron micrograph.

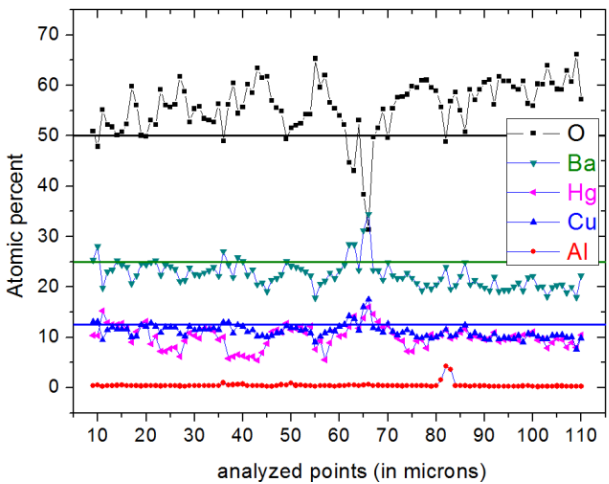
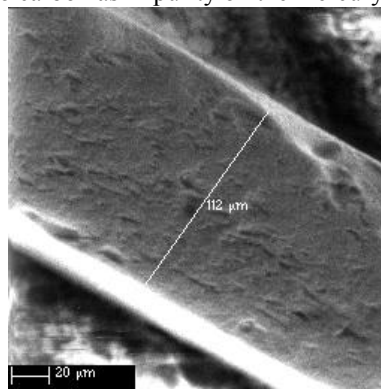

*X-ray Diffraction and Crystal Structure Analysis*: The data were collected at 150(2) K on a Nonius Kappa-CCD area detector diffractometer[1] using graphite-monochromated Mo Kα radiation (λ = 0.71073 Å). The crystals used were small platelets cut from the large synthesized crystals. The unit cell parameters were determined from ten frames, then refined on all data. The data (combinations of φ- and ω-scans giving a complete data set up to θ = 35°, with a minimum redundancy of 10 for 90% of the reflections) were processed with HKL2000.[2] Absorption effects were corrected empirically with the program SCALEPACK.[2] The structures were refined by full-matrix least-squares on $F^2$ with SHELXL-2014.[3] All atoms were refined with anisotropic displacement parameters and no restraint was applied, except for atom O3 which was refined isotropically with restraints on its displacement parameter. The occupancies of atoms Hg1 and O3 was refined, but the small quantity of gold present on the mercury site in the last four samples was not introduced.

***Magnetic measurements:*** Measurements of DC magnetic susceptibility down to 20 K were carried out by using a SQUID magnetometer (Cryogenic Limited S600) in a field of 10 Oe. The field was applied in the *c*-axis direction, perpendicularly to the largest face of the crystalline platelet. The classical zero field cooled (ZFC) procedure was used. The field was applied at the lowest temperature and the ZFC magnetization was recorded as a function of increasing temperatures up to $T > T_c$. The $T_c$ attributed to the different samples has been determined as the temperature corresponding to the middle of the superconducting transition.

***Polarized micro-Raman scattering:*** Raman experiments have been carried out at room temperature using a JY-T64000 spectrometer in single grating configuration fitted with a 1800 grooves/mm grating and a Thorlabs NF533-17 notch filter to block the stray light. The spectrometer was equipped with a nitrogen-cooled back-illuminated CCD detector. The 532 nm excitation line from a diode pump solid state was used. An Olympus microscope with an objective of magnitude ×50 produced a laser spot of a few micrometers. Raman measurements have been performed in backscattering geometry. Crystals were fixed on their edge in such a way as to align the electric field polarizations of the incident and scattered light along the cristallographic *c*-axis (*z* direction). The incident and scattered light wave vectors are parallel to the *a*-axis (*x* direction). In Porto's notation,[4] this corresponds to the *x*(*zz*)*x* configuration. Each Raman spectrum has been obtained in one frame, repeated twice to eliminate cosmic spikes, with an acquisition time of 120 s. The resolution is approximately 1 cm$^{-1}$.

1. Hooft, R.W.W., *COLLECT*, Nonius BV, Delft: The Netherlands 1998.
2. Otwinowski, Z.; Minor, W. Processing of X-Ray Diffraction Data Collected in Oscillation Mode. *Methods Enzymol.* **1997**, *276*, 307–326.
3. Sheldrick, G. M. Crystal Structure Refinement with SHELXL. *Acta Crystallogr., Sect. C* **2015**, *71*, 3–8.
4. Rousseau, D. L.; Bauman, R. P.; Porto, S. P. S. Normal mode determination in crystals, *J. Raman Spectr.*, **1981**, *10*, 253–290.

**Table S1. Atomic Positions in HgBa$_2$CuO$_{4+\delta}$.**

|     | x   | y   | z      | Multiplicity and Wyckoff symbol | Occupancy |
|-----|-----|-----|--------|---------------------------------|-----------|
| Hg1 | 0   | 0   | 0      | 1a                              | < 1       |
| Ba1 | 0.5 | 0.5 | z(Ba1) | 2h                              | 1         |
| Cu  | 0   | 0   | 0.5    | 1b                              | 1         |
| O1  | 0.5 | 0   | 0.5    | 2e                              | 1         |
| O2  | 0   | 0   | z(O2)  | 2g                              | 1         |
| O3  | 0.5 | 0.5 | 0      | 1c                              | << 1      |

**Table S2. Crystal Data, Refined Parameters and Occupancies for pure Hg-1201 Samples.**

|                          | UD55        | UD62        | UD73        | OP93        | OD89        |
|--------------------------|-------------|-------------|-------------|-------------|-------------|
| $T_c$ (K) (mid-transition) | 55          | 62          | 73          | 93          | 89          |
| a (Å)                    | 3.8850(2)   | 3.8799(2)   | 3.8824(1)   | 3.8781(1)   | 3.8759(1)   |
| c (Å)                    | 9.5048(4)   | 9.5084(5)   | 9.5092(4)   | 9.4988(4)   | 9.4796(4)   |
| R1                       | 0.010       | 0.010       | 0.020       | 0.013       | 0.017       |
| wR2                      | 0.025       | 0.021       | 0.052       | 0.032       | 0.036       |
| z(Ba1)                   | 0.30129(3)  | 0.30087(3)  | 0.30039(4)  | 0.29876(3)  | 0.29707(5)  |
| z(O2)                    | 0.2063(4)   | 0.2064(4)   | 0.2072(6)   | 0.2084(5)   | 0.2091(7)   |
| $p^{a)}$                 | 0.09        | 0.096       | 0.109       | 0.16        | 0.183       |
| p/2 (O3 content)         | 0.045       | 0.048       | 0.055       | 0.08        | 0.092       |
| Occupancy(O3)            | 0.007(13)   | 0.020(14)   | 0.08(2)     | 0.11(3)     | 0.16(2)     |
| Occupancy(Hg1)           | 0.9500(16)  | 0.9489(18)  | 0.940(2)    | 0.934(2)    | 0.933(2)    |

**Table S3. Crystal Data, Refined Parameters and Occupancies for gold-substituted Hg-1201 Samples.**

|                          | UD72        | UD88        | OD72        | OD65        |
|--------------------------|-------------|-------------|-------------|-------------|
| $T_c$ (K) (mid-transition) | 72          | 88          | 72          | 65          |
| a (Å)                    | 3.8626(2)   | 3.8721(1)   | 3.8591(1)   | 3.8579(1)   |
| c (Å)                    | 9.4632(3)   | 9.4931(5)   | 9.4649(5)   | 9.4673(4)   |
| R1                       | 0.012       | 0.011       | 0.020       | 0.014       |
| wR2                      | 0.023       | 0.028       | 0.034       | 0.023       |
| z(Ba1)                   | 0.29623(4)  | 0.29896(2)  | 0.29603(4)  | 0.29569(4)  |
| z(O2)                    | 0.2099(5)   | 0.2074(3)   | 0.2108(6)   | 0.2091(5)   |
| $p^{a)}$ | 0.108 | 0.134 | 0.212 | 0.220 |
| $p/2$ | 0.054 | 0.067 | 0.106 | 0.110 |
| Occupancy(Hg1) | 0.9720(12) | 0.9538(15) | 0.9605(16) | 0.9690(15) |

a) Electron hole concentration $p$ in the CuO$_2$ planes of HgBa$_2$CuO$_{4+\delta}$, from Presland, M. R.; Tallon, J. L.; Buckley, R. G.; Liu, R. S.; Flower, N. E. General trends in oxygen stoichiometry effects on $T_c$ in Bi and Tl superconductors. *Physica C* **1991**, *176,* 95–105.